\documentclass[10pt,aps,prb,twocolumn,superscriptaddress,showpacs]{revtex4-2}
\usepackage{amsmath}
\usepackage{amsfonts}
\usepackage{graphicx} 
\usepackage{float}
\usepackage{dcolumn}
\usepackage{bm}
\usepackage{xcolor}
\usepackage{subfigure}
\usepackage{cancel}

\begin{document}

\frenchspacing

\title{Level dynamics and avoided level crossings in driven disordered quantum dots}

\author{Andr\'as Grabarits}
\affiliation{Department of Theoretical Physics, Institute of Physics, 
Budapest University of Technology and Economics, M\H uegyetem rkp. 3., H-1111 Budapest, Hungary}
\affiliation{MTA-BME Quantum Dynamics and Correlations Research Group, 
Budapest University of Technology and Economics,  M\H uegyetem rkp. 3., H-1111 Budapest, Hungary}

\begin{abstract}
The statistical properties of the dynamics of energy levels are investigated in the case of two two-dimensional disordered quantum dot models with nearest neighbor hopping subjected to external time-dependent perturbations. While in the first model the external drivings are realized by a continuous variation of the on-site energies, in the second one it is generated by deformations of a parabolic potential. We concentrate on the effects of the potential on the localization properties and investigate the statistics of the energy level velocities and curvatures regarding their typical magnitudes and domain of agreement with the predictions of Random Matrix Theory (RMT) for the Gaussian Orthogonal, Unitary and Symplectic ensembles. Moreover, the statistical properties of the avoided level crossings are investigated in terms of the corresponding Landau-Zener parameters. We find that the strength of the Landau-Zener transitions exhibit universal behavior which also imply universal single particle dynamics for slow perturbations independent of the disorder and potential strength, the system size and the symmetry class.
These results can be verified experimentally by measurements of single-particle energy spectra in quantum dots.\\

\end{abstract}

\maketitle

\section{Introduction}
The spectra of complex, interacting many-body systems can be considered up to a large extent indeterministic, for which Random Matrix Theory (RMT) has proven to provide an accurate statistical description, relying only on the fundamental symmetries of the system and completely neglecting the microscopical details of the individual energy eigenstates~\cite{Wigner,Bohigas,Mehta,CommonConcept,GuhrCommonconcept}.
 While its applicability to the spectrum of disordered tight binding models has been the subject of many studies~\cite{LStatMITTrans,CoulombBlockSpace,Ando1,Ando2,TwistedBoundaryMetIns,2DGOE}, the statistical behavior of motion of energy levels and its connection to RMT still raises many exciting unanswered questions. Urged by the swift experimental developments exploring non-equilibrium phenomena in the nanoscale regime~\cite{ColdAtoms1,Pekola1,Pekola2}, considerable theoretical attention is being paid to the response properties of disordered quantum dots to time-dependent perturbations. The effects of external drivings manifest themselves, among others, in the movements of energy levels, which in turn provide information about the changes of the physical quantities in the system induced during the non-equilibrium process.

One of the earliest approaches to study disordered systems is provided by the statistics of difference of adjacent energy levels. First of all, as was proposed in Refs.~\cite{Berry,BohigasSinai}, levels repelling each other, characteristic for RMT, correspond to classically chaotic nature and follow the celebrated Wigner Dyson statistics~\cite{Dyson}, while regular motion implies Poissonian statistics. In addition, level spacing statistics also provides an ideal testbed to study localization properties of single particle states. In $3$ dimensions RMT-like behavior is observed for states with localization lengths much larger than the system size and for delocalized states, while Anderson localized states with localization length shorter than the system size exhibit Poissonian level statistics and intermediate statistics were observed  at the metal insulator transition~\cite{MITCritical,MITCritical2,LStatMITTrans,VargaImre1,VargaImre2}, see also a new thorough review on the subject~\cite{Prozen3DAnderson}. A different picture emerges, however, in $2$ dimensions as for spinless disordered tight binding models~\cite{BatschRandomMangetic,RandomMagnetic2,2DGOE,CoulombBlockSpace} RMT-like behavior is only observed for states with sufficiently large ratio of the system size and localization length, $L/\xi$, which is the only relevant parameter according to the single parameter scaling theorem~\cite{Prozen2DAnderson,AndersonLoc,AndersonLoc2,AndersonLoc3,AndersonLoc4,AndersonLoc5}.
Introducing also spin-orbit couplings and considering spin degrees of freedom, truly extended states appear below a critical disorder value~\cite{EvangelouZiman1,Evangelou2,Evangelou3,Ando1,Ando2,AndoZiman}.
 Further works tested the validity of the Wigner statistics in various fields ranging from the early studies of Coloumb blockade~\cite{CoulombBlockSpace} and conductance peak spacings~\cite{ConductancePeak} through the effects of Aharonov-Bohm flux piercing through disordered samples~\cite{BerryRobnik,ABfluxMetal}, kicked one-dimensional systems~\cite{KickedLStat} to the current investigation of interacting spin systems~\cite{Heisenberg1,Heisenberg2,Heisenberg3,MITTrans}, finite range Coulomb gas models~\cite{FRCG1,FRCG2}, and open chaotic systems realised in microwave cavities~\cite{MicroWaveCavity}.\\
Level spacing statistics, however, provides no information about responses to time-dependent perturbations. A large amount of non-equilibrium phenomena in disordered systems induced by external drivings can be addressed by the investigation of the motion of energy levels~\cite{QMChaos}. Level dynamics of classically chaotic systems was first formulated in the pioneering works of Refs.~\cite{Pechuka,Yukawa} exhibiting similar statistical behavior as the spectra of the proper Gaussian random matrix ensembles. With an appropriate parametric evolution of the $H(\lambda)$ disordered Hamiltonian with $\lambda$ promoted to fictitious time, derivatives of energy levels reveal most of the non-equilbrium properties of driven systems. Of central interest are the first and second derivatives commonly referred to as level velocity and level curvature, $v_n=\mathrm dE_n/\mathrm d\lambda,\,K_n=\mathrm d^2E_n/\mathrm d\lambda^2$, respectively, providing information about conductance fluctuations and characterizing the sensitivity of energy levels to changing boundaries in metallic samples~\cite{Edwards,Thouless,VelCurveConduct}.
 While the RMT levels exhibit Gaussian velocity statistics exact curvature distributions resisted evaluation for quite a long time, with many intuitive initial heuristic guesses until exact results were derived for it~\cite{WilkinsonBilliard,CurveStatTail,RMTCuvreTail,Simons1,Simons2,Simons3,ExactCurvature,ExactCurvature2,ExactCurveABflux,FyodorovExactCurve}.
Velocity and curvature statistics constituting an intense area of research have been investigated over the years mostly in systems in the presence of a magnetic field~\cite{CurvatureABflux1,CurvatureABflux2Cond}, in chaotic, irregularly shaped billiards with changing boundaries~\cite{VelStatBilliard,BilliardVelCurve,CurveExperiment,CurveAvCrossChaosBilliard,MicroWavebilliardVelStat,CurveKickedRotorStadiumBilliard}     , and quantum systems with twisted boundary conditions~\cite{BRMTCondCurve,CurveatAndersonLoc,TwistedBoundaryMetIns,LocVelStat,VelCurveTransPoint} or in periodically kicked one-dimensional systems~\cite{CurveTailKickedTop,LocVelStat}, while recent studies considered disordered interacting many-body systems~\cite{HeisenbergMBL,HeisenbergMBL2,HeisenbergMBL3}.

A further striking feature in the course of parametric evolution of disordered systems is the formation of the avoided level crossings. In the seminal work of Wigner~\cite{AvCrossOriginal}, it was argued that the levels of $H(\lambda)$ without any particular symmetries may reach close to each other at some $\lambda_0$ value but finally avoid true crossing points. Apart from providing an ideal testbed to study the degree of emerging chaos in the classical counterparts of quantum systems~\cite{ChaosChem1,ChaosChem2,ChaosChem3,NonIntAvcross,ChaosLZDiracdeltaRegularpart}, avoided level crossings have a dramatic impact on the conditions of adiabatic time evolution in driven disordered systems. Following the common approximate expression around the closest approach of $\lambda_0$,
\begin{equation}
	E_{n+1}(\lambda)-E_n(\lambda)\approx\sqrt{\Delta^2+\gamma^2(\lambda-\lambda_0)^2}\,,
\end{equation}
even for slowly varying $\lambda=\lambda(t)$ driving protocols adiabaticity can be violated via Landau-Zener (LZ) transitions with probability $\exp(-\frac{\pi}{2}\frac{\Delta^2}{\gamma\dot\lambda})$~\cite{LZ1,LZ2} with $\Delta$ and $\gamma$ being the smallest level distance (gap) and asymptotic slope, respectively. While pioneering studies on the LZ parameter statistics and their impact on non-equilibrium dynamics in random matrix ensembles were provided by Wilkinson in Refs.~\cite{WilkAvCrossStat,WilkSingSpectrum}, their strong connection to classical diffusion processes and energy dissipation was established in Refs.~\cite{WilkDiffDissLZ,LZDiffusion,LZResponse,Grabarits1,Grabarits2,Grabarits3}.

The role of the avoided level crossings were studied in further exciting phenomena such as dynamical tunneling~\cite{ChaosTunnelLZ1,ChaosTunnelLZ2,DrivenTunnel3,ResonanceLZ}, relations  to the famous Lyapunov exponents~\cite{LyapunovChaos} or transitions from regular to chaotic regimes in classical systems~\cite{TransChaos1,TransChaos2}. Similarly to level response investigations, LZ parameter statistics were compared to the RMT results in quantum billiards, kicked tops~\cite{OpenBilliardLZ1,OpenBilliardLZ2,AfrciaBilliard,AvCrossStatKickedTopAfricaBiliards,LZDelta_KickedTop} and in disordered systems with Rashba and spin-forbit interactions~\cite{LSAvCross,RashbAvCross}.

In spite of these extensive progresses, in most cases level dynamics were generated by magnetic fields and changing boundaries in microwave cavities or in chaotic billiards not considering the possibility either of spin degrees of freedom or of different driving mechanisms such as deformations of a confining potential which could open new perspectives for experimental realizations. Furthermore, neither LZ parameter statistics nor level dynamics were investigated in two-dimensional systems with random magnetic fields or spin-orbit couplings (GUE, GSE symmetry class, respectively).
In this paper we fill this gap by studying various aspects of level dynamics in two-dimensional disordered tight binding models with different driving protocols feasible for experimental realization and most importantly we point out universal behavior of the LZ parameters implying universal single particle dynamics.\\

The paper is organized as follows. In Sec. \ref{sec:Theoframe} we present the two two-dimensional disordered quantum dot models. In Sec.~\ref{sec:Vel} we investigate the effects of the potential on the localization length and quantify the domain of agreement of the level velocity and curvature statistics with the RMT results and provide analytial resutls on their typical magnitudes. In Sec. \ref{sec:AvCross} we show that the distributions of the Landau-Zener parameters at the anticrossings fall on universal curves identical with the RMT predictions which imply also universal single particle dynamics for slow perturbations and also determine the correspoindg time and parametric velocity units' disorder and size dependences.\\

\section{Theoretical framework}\label{sec:Theoframe}
Before turning to the detailed description of our results we briefly summarize the basic features of RMT. We consider three ensembles of the random matrices, spinless systems with and without time-reversal symmetry described by real symmetric (GOE) and complex hermitian matrices (GUE) and spin one-half systems with time-reversal symmetry with quaternion valued hermitian matrices (GSE), respectively. The distribution of matrix elements for the above classes is given by
\begin{equation}
	\mathcal P\left(H\right)\propto e^{-\frac{\beta N}{4 J^2}\mathrm{Tr}\left(HH^\dagger\right)}
\end{equation}
with $J$ fixing the energy scale and $\beta=1,2,4$ denoting the number of independent real variables of each matrix element for GOE, GUE and GSE, respectively. Neighboring energies exhibit level repulsion as the distribution of their difference in the middle of the spectrum follows the celebrated Wigner-Dyson statistics~\cite{Dyson},
\begin{equation}\label{eq:WD}
	P_{\mathrm{level},\beta}\sim\Delta^\beta e^{-C_\beta\Delta^2}
\end{equation}
with $C_\beta=\frac{\pi}{4}\,,\frac{4}{\pi}\,,\frac{64}{9\pi}$, implying $P_{\mathrm{level},\beta}\left(\Delta\right)\sim\Delta^\beta$ for $\Delta=0$. Following the works ~\cite{WilkNearDegen,WilkSingSpectrum}, we choose the parametric evolution
\begin{equation}\label{eq:rmtprotocol}
	H(\lambda)=H_i\cos\lambda+H_f\sin\lambda
\end{equation}
with $H_i$ and $H_f$  being two independent random matrices drawn from the same ensemble and which ensures identical distribution of matrix elements at any value of $\lambda$.
In particular, we consider the following model on a $L\times L$ square lattice (`on-site model'):
\begin{equation}\label{eq:onsite}
\begin{split}
	H_\text{on-site}(\lambda)=-&J\sum_{\mathbf r,\bm\delta,\alpha,\alpha^\prime} t_{\mathbf{r},\bm\delta,\alpha,\alpha^\prime}\,\lvert\mathbf r+\bm\delta,\alpha\rangle\langle\mathbf r,\alpha^\prime\rvert\\
	&+\sum_{\mathbf r,\alpha}\epsilon_{\mathbf r,\lambda}\,\,\lvert\mathbf r,\alpha\rangle\langle\mathbf r,\alpha\rvert+\,\,\mathrm h.\,\mathrm c.\,,
	\end{split}
\end{equation}
where $J$ sets the energy scale, $\lvert\mathbf r,\alpha\rangle$ and $\lvert\mathbf r+\bm\delta,\alpha\rangle$ denote coordinate and spin eigenstates at lattice site $\mathbf r$ and spin state $\alpha$, respectively, and $\bm\delta$ points to the nearest neighbors. For the spinless case, choosing unit hoppings $t_{\mathbf r,\bm\delta}=1$ GOE-like behavior, while for random phase hopping terms, $t_{\mathbf{r},\bm\delta}=e^{i\varphi_{\mathbf r,\bm\delta}}$, where $\varphi_{\mathbf r,\bm\delta}$ are such that the flux of each plaquette is distributed uniformly in the interval $\left[-\pi,\pi\right]$, statistical properties similar to GUE are expected. For spin one-half systems with time-reversal symmetry, the appropriate choice, $t_{\mathbf r,\bm\delta,\alpha,\alpha^\prime}=\left(V_1\mathbb I+iV_2\sigma_y\right)_{\alpha,\alpha^\prime}$, if $\boldsymbol\delta$ points in the $x$ direction and $t_{\mathbf r,\bm\delta,\alpha,\alpha^\prime}=\left(V_1\mathbb I+iV_2\sigma_x\right)_{\alpha,\alpha^\prime}$ for $\boldsymbol\delta$ parallel to the $y$ axis, leads to an energy spectrum characteristic for the GSE ensemble. Here $\sigma_x$ and $\sigma_y$ denote the $x$ and $y$ Pauli matrices and following the convention of Ref.~\cite{Ando2} we chose $V_1=\sqrt3/2,\,V_2=1/2$ for the strength of the spin-orbit coupling. Evolution in parameter space is realized by the protocol $\epsilon_{\mathbf 
r,\lambda}=\epsilon_{\mathbf r,i}\cos(\lambda)+\epsilon_{\mathbf r,f}\sin(\lambda)$ starting from $\lambda_i=0$ and ending at $\lambda_f=\pi/2$, where $\epsilon_{\mathbf r,i}$ and $\epsilon_{\mathbf r,f}$ are independent and distributed uniformly on $[-W/2,W/2]$. Note that this is the same parametric evolution as in the case of the RMT protocol \eqref{eq:rmtprotocol} but now with randomness only encoded in the on-site energies and hopping terms making it more feasible for experimental realizations.

In the case of the `potential model' we consider the following Hamiltonian:
\begin{equation}\label{eq:pot}
\begin{split}
	H_\text{pot}(\lambda)=-&J\sum_{\mathbf r,\bm\delta,\alpha,\alpha^\prime}t_{\mathbf{r},\bm\delta,\alpha,\alpha^\prime}\,\lvert
	\mathbf r+\bm\delta,\alpha\rangle\langle\mathbf r,\alpha^\prime\rvert\\
	 &+\sum_{\mathbf r,\alpha}\left(V_{\mathbf r,\lambda}+\epsilon_{\mathbf r}\right)\lvert\mathbf r,\alpha\rangle\langle
	 \mathbf r,\alpha\rvert+\,\,\mathrm h.\,\mathrm c.\,
	\end{split}
\end{equation}
where level dynamics is now generated by the compression (decompression) of a parabolic potential, $V_{\mathbf r,\lambda}=\frac{1}{2}\frac{V_0}{L^2}\left(r^2+\lambda\left(x^2-y^2\right)\right)$,  in the $x$ $(y)$ direction in a symmetric way with $\lambda$ starting at $-\lambda_f/2$ and ending at $\lambda_f/2$.
Here, choosing the same $t_{\mathbf{r},\bm\delta,\alpha,\alpha^\prime}$ hopping terms and on-site energies as in the on-site model~\eqref{eq:onsite}, statistical properties of level dynamics are expected to be identical to the proper random matrix ensembles (i.e. unit, random phase and SU(2) phase hoppings implying GOE, GUE and GSE-like behavior, respectively), albeit under slightly different conditions due to the presence of the confining potential.

\section{Level dynamics}\label{sec:Vel}
This section is devoted to the analysis of the localization properties and the statistics of the level velocities and curvatures in the two quantum dot models.

\subsection{Effects of the potential on the localization properties}
In this subsection we deal with the analysis of the localization properties of the potential model captured by the level spacing statistics.
While without the potential term the localization length depends only on the disorder strength and the position in the energy spectrum according to the single parameter scaling theorem, $\xi=\xi(W,E)$ ~\cite{AndersonLoc,AndersonLoc2,AndersonLoc3,AndersonLoc4,AndersonLoc5}, the presence of the quadratic potential slightly modifies this picture.
First it stretches the spectrum upwards and shifts the zero energy states towards the lower band edge, second, changing the on-site energies it also exhibits a non-trivial interplay with the disorder strength. Due to the scaling $V_0/L^2$ in~(\ref{eq:pot}), however, the on-site energy contributions of the potential are independent of the system size.  Hence, it is expected that for fixed potential strength the single parameter scaling remains valid with a modified, potential dependent localization length, $\xi(W,E)\rightarrow\xi_V(W,E,V_0)$, under the numerically observed condition that the lowest energy contributions in the middle of the sample do not exceed extremely the energy scale, $V_0/(2L^2)<J$.
Although the precise analysis of the localization length is beyond the goals of this paper, the essential characteristics can be captured in terms of the average level spacing ratio, $r_n=\frac{\mathrm{min}\left\{\delta_n,\,\delta_{n-1}\right\}}{\mathrm{max}\left\{\delta_n,\,\delta_{n-1}\right\}}$ with $\delta_n=\varepsilon_{n+1}-\varepsilon_n$ denoting the spacing between the $n+1$th and $n$th levels. In particular, using the results on the disorder dependences of the localization lengths of Refs.~\cite{2DGOE,RandomMagnetic2,Ando2}, we identified the new $\xi_V$ localization lengths with the ones in the potential free models at which the level spacing ratios matched. 

\begin{figure}[t]
\includegraphics[width=0.48\textwidth]{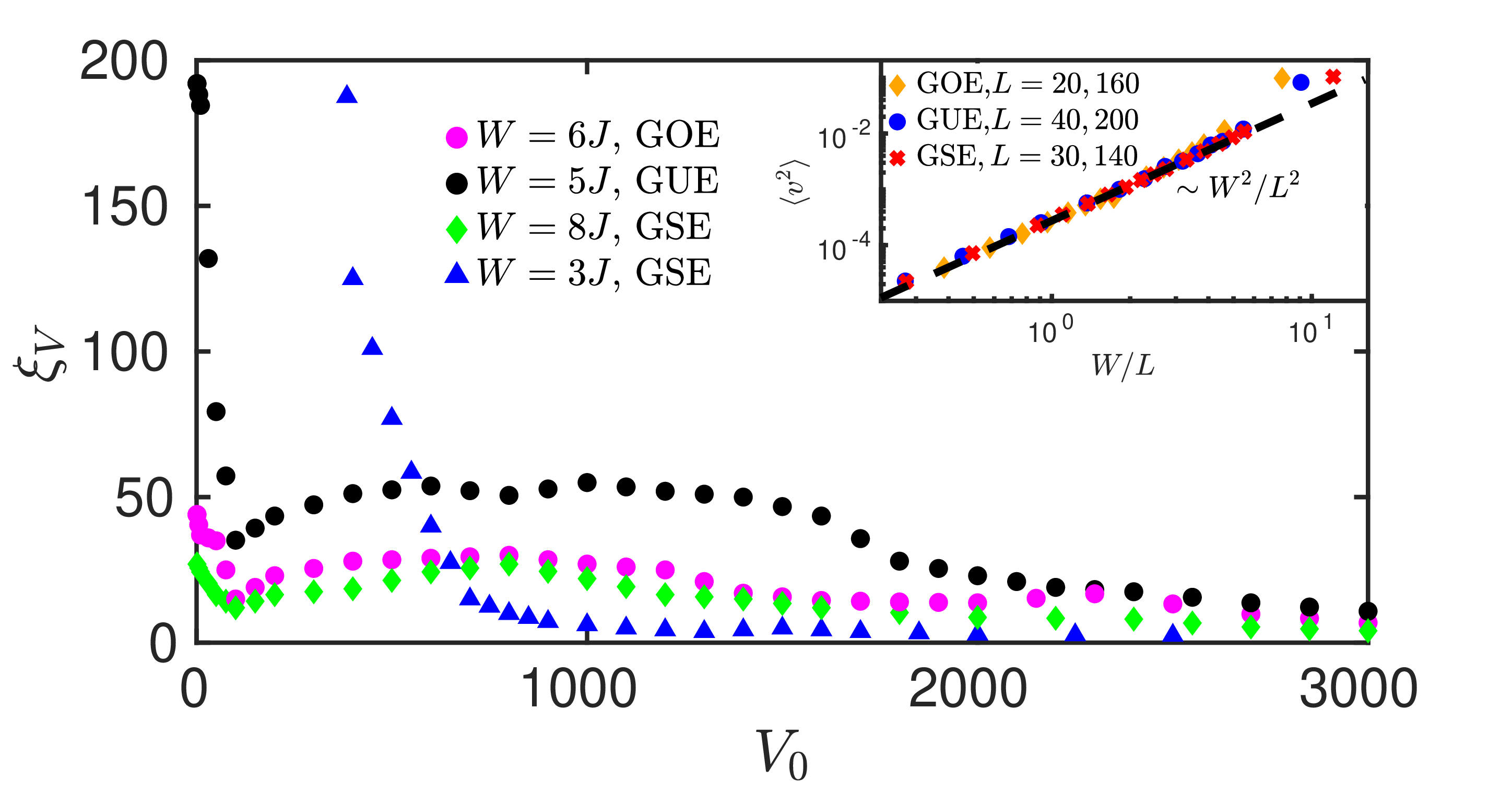}
\caption{Localization length as a function of the potential strength for various system sizes and disorder strengths. For weak potentials they agree with the ones in the corresponding Anderson models and for the GSE ensembles even for $W=3<W_c$ finite localization length is observed for $V_0>300J$. Inset: Velocity variance scaling as $\sim W^2/L^2$ with the same prefactor for the three ensembles.}
\label{fig:Loc_V}
\end{figure}

For fixed potential strengths below the threshold, $V_0<2L^2J$, up to numerical precision the obtained localization lengths depended only on the disorder strength checked for various system sizes ranging from $L=20$ to $L=200$ confirming our assumption about the validity of the single parameter scaling theorem.\\
The obtained dependences of the localization lengths on the potential strength are plotted on Fig.~\ref{fig:Loc_V} for fixed disorder strengths in the small, intermediate and strong localizaton regimes for all the three ensembles for various system sizes ranging from $L=20$ to $L=200$. For small potential strengths, $V_0\sim J$, the localization lengths agree with those in Refs.~\cite{2DGOE,RandomMagnetic2,Ando2}, while they decrease with increasing $V_0$.
Furthermore, in the GSE ensemble the confining potential decreases the value of the critical disorder, $W_c\approx 5.875J$ observed in the finiteness of $\xi_V(W,E,V_0)$. Without precise analysis (being, however, an interesting future direction) it is demonstrated for $W=3$ with the critical potential strength  $V^c_0\approx300J$ accurate up to system size $L=200$.

\subsection{Level velocity statistics}
Next we discuss the statistical properties of the energy level velocities both in the RMT and the strongly localized regime. The velocity can be expressed as the first derivative with respect to the parameter $\lambda$:
\begin{equation}\label{VelDef}
\begin{split}
	&v_n\equiv\frac{\mathrm d E_n}{\mathrm d\lambda}=\big\langle\varphi_{n,\lambda}\big\lvert\mathrm dH
	\left(\lambda\right)/\mathrm d\lambda\big\rvert\varphi_{n,\lambda}\big\rangle\\
	&=\sum_\mathbf{r}\,\lvert\varphi_{n,\lambda}(\mathbf r)\rvert^2\,\partial_\lambda\varepsilon_{\mathbf r,\lambda}
	\end{split}
\end{equation}
with the shorthand notation $\partial_\lambda\varepsilon_{\mathbf r,\lambda}=-\varepsilon_{\mathbf r,i}\sin\lambda+\varepsilon_{\mathbf r,f}\cos\lambda$ having variance $W^2/12$ and with $\varphi_{n,\lambda}(\mathbf r)=\langle\mathbf r\vert\varphi_{n,\lambda}\rangle$ with $\langle\mathbf r\rvert$ denoting the $\mathbf r$th coordinate eigenstate at parameter value $\lambda$. In the RMT protocol, \eqref{eq:rmtprotocol}, it exhibits a Gaussian distribution, $\mathcal P\left(v_n\right)\sim e^{-\frac{\beta N}{4} v^2_n}$, independently of the $\lambda$ parameter.

In the on-site model, \eqref{eq:onsite} first we determined the domain of validity of the RMT description quantified by the integral of the absolute value difference of the numerically obtained statistics with variances scaled to unity from the Gaussian curve, $\int_{-\infty}^\infty\mathrm dv\,\left\lvert\mathcal P(v)-\exp\left(v^2/2\right)/\sqrt{2\pi}\right\rvert$. We fixed the deviation to be $0.1$ and numerically found the corresponding threshold ratios, $\xi/L\approx0.77,\,0.4,\,0.34$ for the GOE, GUE and GSE classes, respectively, checked for system sizes and disorder strengths $L=20,\dots,200$, $W=1J,\dots,10J$, respectively.

As far as the weakly and strongly localized limits are concerned, in the Gaussian regime the eigenstate components are uniformly distributed on a $\beta L^2$ dimensional sphere with absolute value squares becoming independent identically distributed random variables with mean $1/L^2$ in the limit $L\gg1$. Consequently, level velocity in \eqref{VelDef} is given by a sum of independent identically distributed random variables.
Since $\partial_\lambda\varepsilon_{\mathbf r,\lambda}$ is independent of $\varepsilon_{\mathbf r,\lambda}$ and thus of $\lvert\varphi_{n,\lambda}(\mathbf r)\rvert^2$ as well the total variance is the sum of each term's variance, which by the Central Limit Theorem results in a total variance of $\sim W^2/L^2$ (see the inset of Fig.~\ref{fig:Loc_V}) and a Gaussian velocity statistics independent of the underlying symmetry class.

In the strongly localized regime, the velocity statistics becomes the same as those of the Hamiltonian's matrix elements with variance scaling as $\sim W^2$, as the sum \eqref{VelDef} results in just one term, where the eigenstates are localized with exponential accuracy.

Turning to the potential model, ~\eqref{eq:pot}, according to the previous subsection, with the modified $\xi_V(W,E,V_0)$ localization lengths, the same threshold of the absolute value deviation from Gaussian is reached around the values $L/\xi_V\approx 0.41,\,0.29,\,0.19$ for the GOE, GUE and GSE classes, respectively. Note that due to the different protocols the thresholds are slightly smaller than in the on-site model.
In contrast to the RMT and on-site results, the mean of the velocity can also take finite values growing linearly with $\lambda V_0/L^2$ with the observed prefactor of $0.6$. Similarly, other statistical properties also acquire additional subleading $\sim\lambda V_0/L^2$ correction terms, however, for $\lambda<1$ their effects are negligible compared to the $\lambda=0$ point. Considering next the statistical behavior in the weakly and strongly localized limits first the velocity is expressed similarly to Eq.~\eqref{VelDef}:
\begin{equation}\label{eq:VelPot}
	v_n=\frac{V_0}{2L^2}\sum_\mathbf{r}\,\lvert\varphi_{n,\lambda}(\mathbf r)\rvert^2\,(x^2-y^2)\,,
\end{equation}
where only $\lvert\varphi_{n,\lambda}(\mathbf r)\rvert^2$ is of statistical nature which, however, does not depend on the on-site variances in the RMT regime, implying no disorder dependence of the variance. Investigating further Eq.~\eqref{eq:VelPot}, one can deduce that its variance does not depend on the system size up to leading order either. In its expression, 
\begin{equation}
\begin{split}
	&\langle v^2_n\rangle\sim L^{-4}\sum_{\mathbf r}\langle\lvert\varphi_{n,\lambda}(\mathbf r)\rvert^4\rangle(x^2-y^2)^2\\
	&+L^{-4}\sum_{\mathbf r\neq\mathbf r^\prime}\langle\lvert\varphi_{n,\lambda}(\mathbf r)\rvert^2\lvert
	\varphi_{n,\lambda}(\mathbf r^\prime)\rvert^2\rangle((x^\prime)^2-(y^\prime)^2)(x^2-y^2)\,.
	\end{split}
\end{equation}
$(x^2-y^2)$ and the $\sum_{\mathbf r}$ summations give $\sim L^2$ contributions while the two averages scale as $\sim L^{-4}$. The first term then gives $J^2L^{-4}L^{-4}L^6\sim J^2L^{-2}$ and the second one $\sim\mathrm{O}(J^2)$ in the RMT regime. Additionally, here all terms' variances and correlations are negligibly small compared to the total variance validating the applicability of the CLT as $\langle\lvert\varphi_{n,\lambda}(\mathbf r)\rvert^2\lvert\varphi_{n,\lambda}(\mathbf r^\prime)\rvert^2\rangle(x^2-y^2)((x^\prime)^2-(y^\prime)^2)/L^4<L^{-2}L^{-2}L^2L^2L^{-4}<L^{-4}$. This implies Gaussian distribution with $\mathrm O(J^2)$ variance also independent of the underlying symmetry class with a subleading $\sim 0.8\lambda^2V^2_0/L^4$ correction term observed numerically. In the strongly localized regime, however, level velocities exhibit completely featureless statistics as randomness is now encoded in the $\mathbf r$ positions where the $n$th eigenstate is localized carrying a contribution of $\frac{V^2_0}{2L^2}(x^2-y^2)$. Note that, neither of the above strongly localized limit results match the exact formula derived in Ref.~\cite{LocVelStat}.
The results are demonstrated in Fig.~\ref{fig:VEL} with the statistics normalized such that their maximum values are fixed at $1$, $\tilde{\mathcal P}(\tilde v)=\mathcal P(\tilde v)/\mathrm{max}\left\{\mathcal P(\tilde v)\right\}$ and scaled to have unit variances, $\tilde v=v/\sqrt{\langle v^2\rangle}$ with the potential model results plotted for the $L^2/2$th state for $V_0=100J$ and the $L^2/40$th state for $V_0=1500J$with $L=140$, where in the latter case the zero energy states has been shifted reasonably.
In total, the variances read:
\begin{align}
	&\langle v^2\rangle_\mathrm{on-site}\sim W^2/L^2,\,\,\,\,\text{ independent of }\beta\,,\\
	&\langle v^2\rangle_\mathrm{pot}=\mathrm O(J^2)+\mathrm o(L^{-2})+\mathrm o(\lambda V_0/L^2)\,,\\
	&\text{independent of }W\text{ and }\beta\nonumber\,.
\end{align}

\subsection{Level curvature statistics}

Turning to the curvature of energy levels, characterized by the second derivative with respect to $\lambda$, a compact expression is provided by second order perturbation theory:
\begin{equation}\label{CurveDef}
	\begin{split}
	&K_n\equiv\frac{\mathrm d^2E_{n,\lambda}}{\mathrm d\lambda^2}\\
	&=\left\langle\varphi_{n,\lambda}\left\lvert\mathrm d^2H/\mathrm d\lambda^2\right\rvert\varphi_{n,\lambda}\right\rangle+2\sum_{m\neq n}\frac{\left\lvert \left\langle
	\varphi_{n,\lambda}\left\lvert \mathrm d H/\mathrm d\lambda\right\rvert\varphi_{m,\lambda}\right\rangle\right\rvert^2}
	{E_{n,\lambda}-E_{m,\lambda}}
	\end{split}
\end{equation}
with $\lvert\varphi_{n,\lambda}\rangle\,$ and $E_{n,\lambda}$ denoting again the instantaneous eigenstates (eigenvalues) and being independent of $\lambda$ for RMT~\eqref{eq:rmtprotocol} and the on-site protocols~\eqref{eq:onsite}. In the potential model, we concentrate again on the $\lambda=0$ point.

The RMT result derived in Refs.~\cite{ExactCurvature,ExactCurvature2,ExactCurveABflux} reads
\begin{equation}\label{eqCurve}
	P(K)=\frac{C_\beta(W,L,V_0)}{\left(1+K^2/\gamma^2(W,L,V_0)\right)^{\beta/2+1}}
\end{equation}
with $C_\beta$ and $\gamma$ being the normalization constant and the natural unit of the curvature, respectively, by which the distributions with different parameters fall on identical curve for each ensemble.

Following the strategy of Refs.~\cite{ExactCurvature,ExactCurvature2}, we restricted the investigation to levels around zero energy and computed numerically the second derivative of the energy levels being closest to zero at $\lambda=0$. While in the on-site model,~\eqref{eq:onsite} and RMT the first term in \eqref{CurveDef} becomes $-E_{n,0}$ and so can be neglected, in the potential model \eqref{eq:pot} it is exactly zero as $\mathrm d^2H_\mathrm{pot}/\mathrm d\lambda^2=0$.
Again the threshold value of the $L/\xi$ ratio was found numerically, above which the absolute value deviation from the expression in Eq.~\eqref{eqCurve} gets bigger than $0.1$, giving relatively small values of $L/\xi\approx0.07,0.16,\,0.2\,$ for the three ensembles.
The possible reason is that in the expression of the curvature states far away from $E=0$ also contribute significantly.
In the potential model, similarly to the velocity statistics, slightly different values are observed, $L/\xi_V\approx0.171,\,0.09,\,0.07,\,$ which is again due to the different underlying mechanism governing the parametric evolution.

Although in the RMT protocol the unit scales as $\gamma^2(W,L,V_0)\sim\beta\langle v^2\rangle/\delta\epsilon$~\cite{ExactCurvature,ExactCurvature2} with $C_\beta$ depending only on the symmetry class and $\delta\epsilon$ denoting the mean level spacing at $E=0$, different behavior is observed in the investigated models. These differences are again, most probably, due to the localized nature of the eigenstates far from $E=0$.

\begin{figure}[t]
\includegraphics[width=0.48\textwidth]{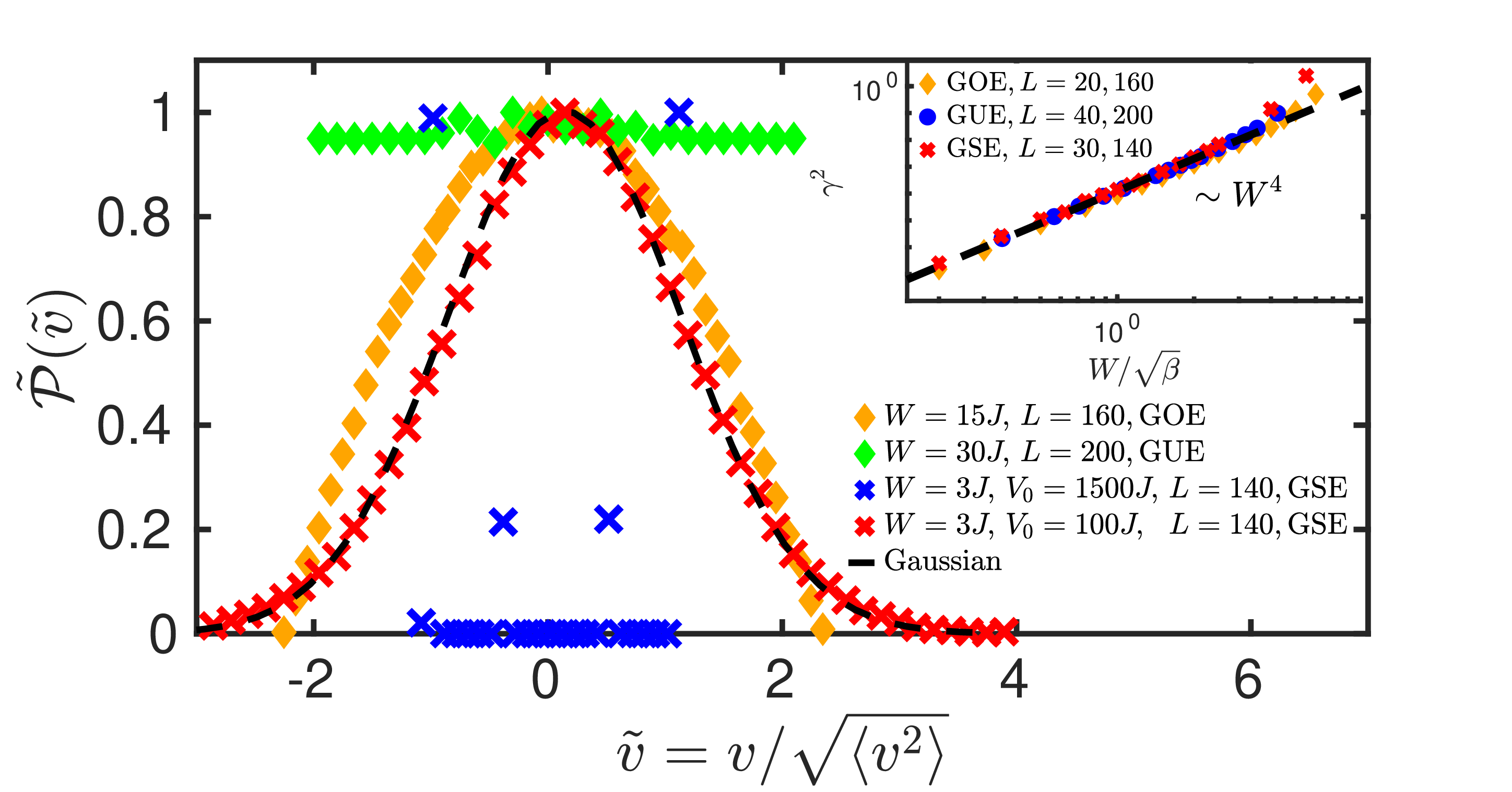}
\caption{Velocity distributions normalized such that their maximum values are $1$  with variances scaled to unity. In the on-site model the statistics transforms from Gaussian to uniform distribution as disorder increases (diamonds), while for large potential strength completely featureless statistics is observed (crosses) in the potential model. Inset: Curvature unit dependence in the on-site model for the three ensembles, exhibiting $\beta^2W^4$ power-law behavior.}
\label{fig:VEL}
\end{figure}

 To this end we numerically investigated the size and disorder dependence of the curvature unit and found that in the on-site model, quite unexpectedly, it scales as  $\gamma^2\sim W^4/J^2$ (see the inset of Fig.~\ref{fig:VEL}), while it does not depend on the system size in the RMT-like regime as the numerator scales similarly to the velocity which together with the denominator's typical magnitude of $\sim L^{-2}$ gives a scaling of $\mathrm O(J^2)$.

In the potential model writing out the numerator, $\left\langle
	\varphi_{n,\lambda}\left\lvert \mathrm d H/\mathrm d\lambda\right\rvert\varphi_{m,\lambda}\right\rangle\sim L^{-2}\sum_{\mathbf r}\varphi^*_{n,\lambda}(\mathbf r)\varphi_{m,\lambda}(\mathbf r)(x^2-y^2)$, where in contrast to Eq.~\eqref{eq:VelPot} the summation only yields a factor of $L$ due to the fluctuating phase in $\varphi^*_{n,\lambda}(\mathbf r)\varphi_{m,\lambda}(\mathbf r)$ leading to a typical magnitude of $\left\lvert\left\langle
	\varphi_{n,\lambda}\left\lvert \mathrm d H/\mathrm d\lambda\right\rvert\varphi_{m,\lambda}\right\rangle\right\rvert^2$ scaling as $\sim L^{-2}$, which together with the denominator's similar $\sim L^{-2}$ scale results in $\mathrm O(J^2)$.
 In both models, moreover, $\gamma^2\sim\beta^2$ is observed as the number of terms in the numerator increases linearly with $\beta$.
Summarizing:
\begin{align}
	&\gamma^2(W,L)_\mathrm{on-site}\sim\beta^2W^4/J^2\,,\text{ independent of }L\,,\\
	&\gamma^2(W,L,V_0)_\mathrm{pot}\sim\beta^2O(J^2)+\mathrm o(\lambda V_0/L^2)\,,\\
	&\text{independent of }W\text{ and }L\nonumber\,.
\end{align}
Finally, in the strongly localized limit, a quite simple feature emerges, where the numerator simply disappears in Eq.~\eqref{CurveDef} implying that levels cross the zero-energy point as straight lines with zero higher than first order derivatives.

\section{Statistics of avoided level crossings}\label{sec:AvCross}
In this section, having summarized the essential characteristics of the statistical behavior of the parametric evolution of energy levels, we turn to the main message of this paper, the characterization of avoided level crossings and the aspects of the strongly related universal single particle dynamics.
\subsection{Distribution of the Landau-Zener parameters}
In this section we first investigate the statistics of the LZ parameters, i.e. the gap and asymptotic slope of energy levels at the avoided level crossings which determine the strength of level-to-level transitions in slow parametric time-evolutions. In systems with no particular symmetries single parameter variations in random Hamiltonians induce avoided crossings, neighboring levels approaching very closely to each other, but finally avoiding true degenerate points. Such an anticrossing, located at $\lambda_0$, can be described by the effective $2\times 2$ matrix, in the limit that the separation from the other levels are much larger than the typical distance between the two energies:
\begin{align}\label{eq:LZ}
	H&=\begin{bmatrix}\lambda\gamma/2&\Delta_\mathrm{min}\\\Delta_\mathrm{min}&-\lambda\gamma/2\end{bmatrix},\\
	\Delta\left(\lambda\right)&=E_+\left(\lambda\right)-E_-\left(\lambda\right)=\sqrt{\Delta^2_\mathrm{min}+
	\gamma^2\left(\lambda-\lambda_0\right)^2}
\end{align}
with $E_{\pm}$, $\Delta_\mathrm{min}$ and $\gamma$ denoting the two eigenvalues, the gap and the asymptotic slope, respectively.\\
In the case of random matrices, concentrating on the middle of the energy spectrum, statistics of $\Delta$ and $\gamma$ were calculated by Wilkinson in Refs.~\cite{WilkSingSpectrum,WilkAvCrossStat} for the protocol ~\eqref{eq:rmtprotocol} when the value of $\lambda$ is changed from $0$ to $\pi/2$:
\begin{align}
	&\rho_\beta\left(\tilde\Delta_\mathrm{min}\right)\sim \tilde\Delta^{\beta-1}e^{-C_{\beta,\Delta}\tilde\Delta^2_\mathrm{min}}\,,\\
	&\rho_\beta\left(\tilde\gamma\right)\sim \tilde\gamma^{\beta+1}e^{-C_{\beta,\gamma}\tilde\gamma^2}\,,
\end{align}
with $C_{\beta,\Delta}=\frac{1}{\sqrt\pi},\,\frac{\pi}{4},\,\frac{9\pi}{16}$ and $C_{\beta,\gamma}=\frac{4}{\pi},\,\frac{9\pi}{64},\,\frac{225\pi}{256}$ for GOE, GUE and GSE, respectively and where the following dimensionless quantities were introduced $\tilde\Delta_\mathrm{min}\equiv\Delta_\mathrm{min}/\langle\Delta_\mathrm{min}\rangle$, $\tilde\gamma\equiv\gamma/\langle\gamma\rangle$ chosen such that the statistics have unit mean. Note that for both cases the distributions are independent and follow different curves for the symmetry classes $\beta=1,2,4$, respectively.\\

\begin{figure*}[t!]
\includegraphics[width=.85\textwidth]{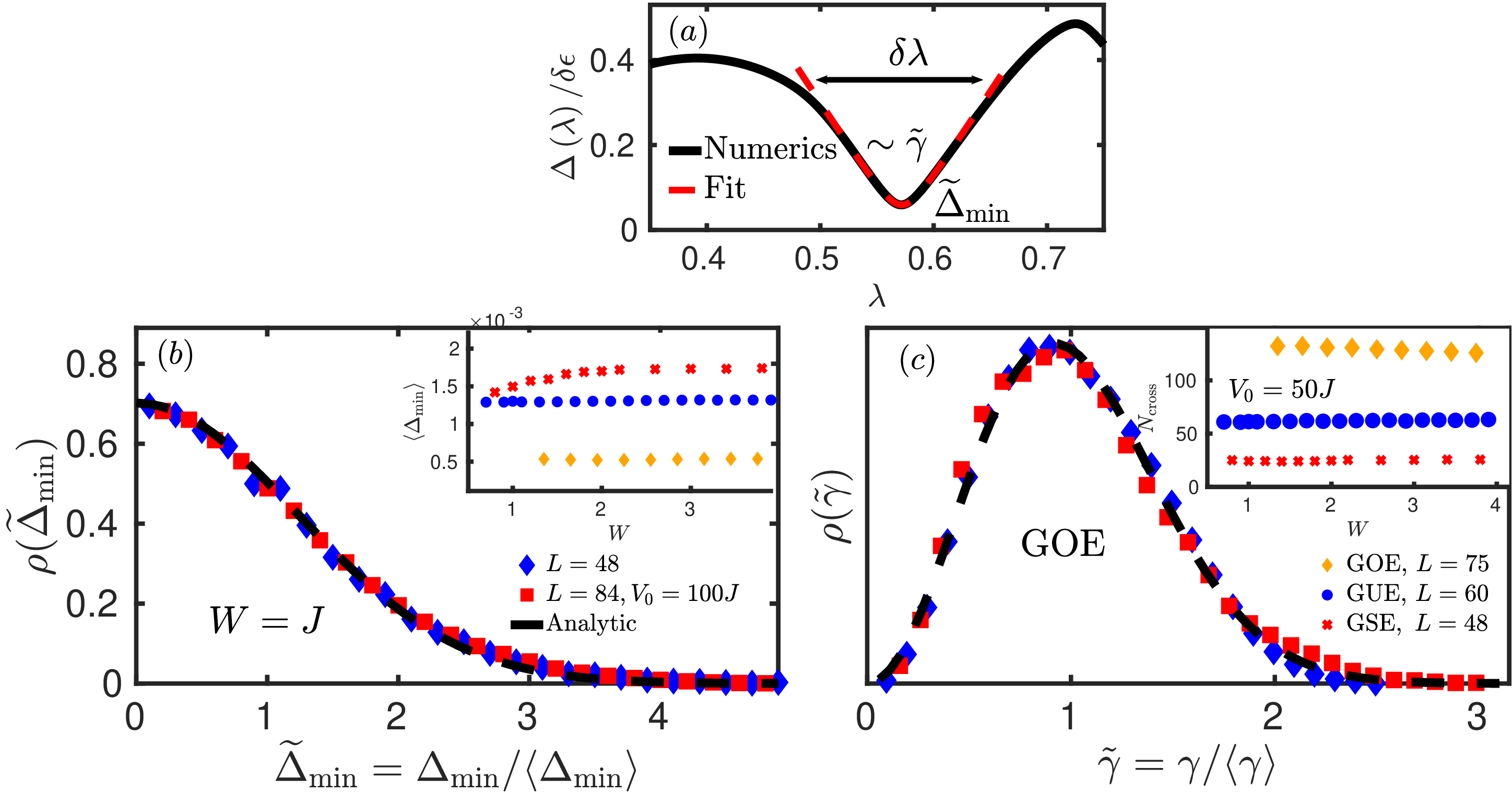}
\caption{Comparison of the distribution of the gap and the asymptotic slope at the avoided level crossings obtained in the two-dimensional models and the RMT analytical results for the GOE ensemble (symbols). (a): Typical shape of an avoided crossing with width $\delta\lambda$, slope $\tilde\gamma$ and gap $\tilde\Delta_\mathrm{min}$, with the dashed line fitting the Landau-Zener approximation of the level distance. (b): Gap distribution for the on-site and potential model. Red squares: potential model for system size, variance and potential strength, $L=28$, $W=J,$ and $V_0=100J$, respectively. Blue diamonds: on-site model results for $L=24$ and $W=J$. Good agreement is observed between the distributions with their means scaled to unity and the analytical results (dashed line). Inset: Disorder dependence of the gap in the potential model for $V_0=50J$ showing insensitivity, up to numerical precision, for all the three ensembles.
(c): Distributions of the asymptotic slope for the same parameters. Numerical data collected from around zero energy and scaled to have unit mean values are in good agreement with the RMT analytical results (dashed line). Inset: Disorder dependence of the number of the avoided level crossings in the potential model for $V_0=50J$, exhibiting again constant behavior up to numerical precision.}
\label{GOELZ}
\end{figure*}

To this end we collected numerical data of the Landau-Zener parameters for several disorder realizations between the  levels around the zero energy states (i.e. the middle of the spectrum in the on-site model and RMT) and compared the obtained statistics scaled to unit mean values to the RMT results. As far as their sensitivity to localization is concerned, while gap statistics behave similarly to level spacing statistics, slope statistics show patterns similar to level velocity statistics.
In particular, we numerically determined again the threshold ratios above which the integral of the absolute value deviation of the numerically obtained statistics from the analytical ones exceeds $0.1$ yielding $L/\xi\approx0.77,\,0.4,\,0.34$, $L/\xi_V\approx 0.41,\,0.29,\,0.19$ and $L/\xi\approx L/\xi_V\approx0.75,0.48,0.19$ for the slope and gap, respectively.
Turning to the investigation of the typical magnitudes of the LZ parameters first note that both parameters can only take positive values so it is sufficient to consider their mean values instead of their variances. Similarly to the threshold ratios, the asymptotic slope and gap behave similarly to the velocity and level spacing, respectively.
 In agreement with this latter statement, slope scales with the disorder strength and the system size exactly the same way as it was observed for the level velocities, i.e. in the on-site model $\langle\gamma\rangle\sim WL^{-1}$, while it is independent of both parameters in the case of the potential model and it does not depend on the particular ensemble either up to leading order. 
 Regarding further the typical magnitude of the gaps we observe that it scales with the system size as $\langle\Delta_\mathrm{min}\rangle\sim JL^{-2}$ in both models. Although it is the expected scale, it does not match the general result for the off-diagonal matrix element, $\left\langle\varphi_{n,\lambda}\left\lvert \mathrm d H/\mathrm d\lambda\right\rvert\varphi_{m,\lambda}\right\rangle\sim L^{-1}$.
Moreover, quite surprisingly it is insensitive to the disorder strength up to numerical precision. Remarkably it also implies that, in strong contrast to RMT, the gap is not proportional to the mean level spacing (for RMT-like states) which latter does increase with the disorder strength. 

\begin{figure*}[t]
\includegraphics[width=.77\textwidth]{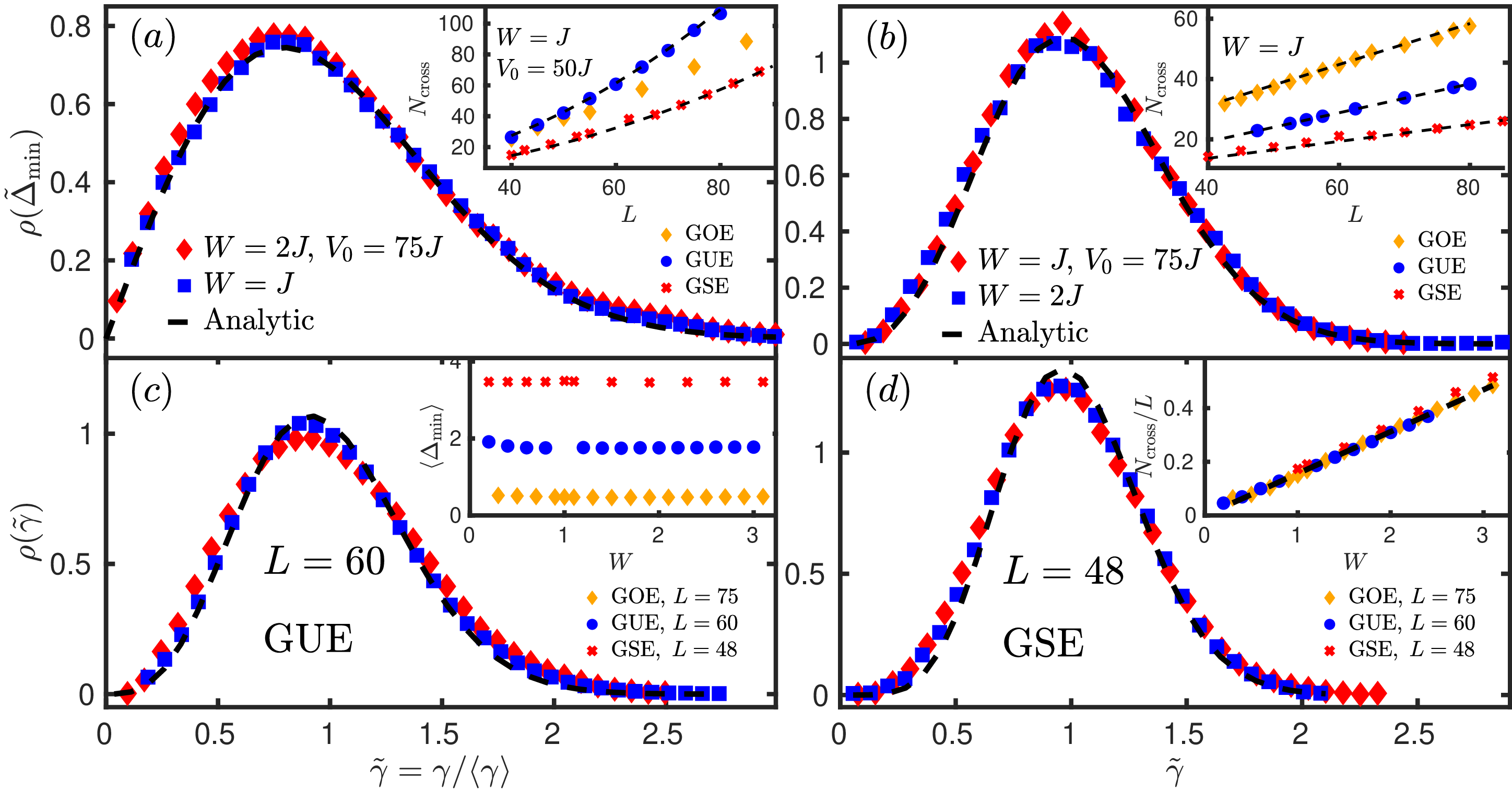}
\centering
\caption{Statistics of the Landau-Zener parameters at the avoided crossings for the GUE and GSE ensembles comparing the RMT numerical results with the on-site and potential model (symbols). (a): Gap distribution for the GUE ensemble for system sizes, disorder strengths and potential strengths, $L=60$, $W=2J,\,J$ and $V_0=75J$ for the potential model and on-site model, respectively. Inset: Average number of the avoided level crossings in the potential model with $V_0=50J$ and $W=J$ as a function of the system size growing as $\sim L^2$ (dashed line) for all the three ensembles (b): Gap distribution for the GSE ensemble for parameters $L=48$, $W=J,\,2J$ and $V_0=75J$, respectively for the potential and the on-site model. In both cases remarkable agreement is observed with the RMT analytical results (dashed line). Inset: Average number of the anticrossings as function of the system size at $W=J$ for the on-site model growing linearly (dashed line).
(c) and (d): Slope distributions for the same parameters, with the two disordered models following the same curves up to high precision (dashed line). Inset of (c): Scaling of the gap as a function of the disorder strength in the on-site model showing constant behavior up to numerical precision. Inset of (d): Average number of the avoided crossings scaled with the size of the system for the on-site model growing linearly with the disorder strength (dashed line) and with approximately the same coefficient for all the three ensembles. }
\label{GUSELZ}
\end{figure*}

Next we turn to the analysis of the typical spacing, $\Delta\lambda$, between adjacent anticrossings and the typical width, $\delta\lambda$, of them, i.e. the approximate region where the formula Eq.~\eqref{eq:LZ} holds up to good precision. As pointed out in Refs.~\cite{WilkAvCrossStat,WilkSingSpectrum,WilkDiffDissLZ,LZDiffusion,LZResponse,Grabarits1,Grabarits2,Grabarits3}, in slowly driven disordered systems, where RMT description is applicable for the instantaneous energy spectrum, dynamics is very well captured by classical diffusion of hardcore particles in energy space, where transitions happen at the avoided level crossings, approximately around the $\delta\lambda$ region of the closest approach. Furthermore, of utmost importance is its relation to the typical spacing between adjacent avoided crossings providing information about the geometrical structure or "typical shape" of the anticrossings.
While, comprehensively, the typical width should scale as the ratio of the gap and the slope,
\begin{equation}
	\delta\lambda\sim\frac{\langle\Delta_\mathrm{min}\rangle}{\langle\gamma\rangle}\,,
\end{equation}
the typical spacing is captured by counting the average number of the avoided crossings, $N_\mathrm{cross}$, being inversely proportional to the spacing, $\Delta\lambda\sim N^{-1}_\mathrm{cross}$. In RMT we previously showed~\cite{Grabarits1} that the average number grows as $N_\mathrm{cross}\sim\sqrt N$. Counting also the number of the anticrossings while analyzing their LZ parameters we found that in the on-site model it also grows with the square root of the number of levels, $N_\mathrm{cross}\sim L$,  while in the potential model it increases linearly with the number of lattice sites, $N_\mathrm{cross}\sim L^2$. As far as the disorder dependence is concerned, our numerical results show that, as one would expect, up to high precision insensitivity to disorder is observed in the potential model, while in the on-site model it grows as $N_\mathrm{cross}\sim WJ^{-1}$. So in total we obtain for the parameters:
\begin{align}
	&\Delta\lambda_\mathrm{pot}\sim L^{-2},\,\,\,\,\,\,\,\,\,\,\,\,\,\,\,\,\,\,\,\Delta\lambda_\mathrm{on-site}
	\sim JW^{-1}L^{-1}\,,\\
	&\langle\Delta_\mathrm{min}\rangle_\mathrm{pot}\sim JL^{-2},\,\,\,\,\,\langle\Delta_\mathrm{min}\rangle_
	\mathrm{on-site}\sim JL^{-2}\,,\\
	&\langle\gamma\rangle_\mathrm{pot}\sim\mathrm O(J),\,\,\,\,\,\,\,\,\,\,\,\,\,\,\,\,\langle\gamma\rangle_
	\mathrm{on-site}\sim WL^{-1}\,,\\
	&\delta\lambda_\mathrm{pot}\sim L^{-2},\,\,\,\,\,\,\,\,\,\,\,\,\,\,\,\,\,\,\,\,\,\delta\lambda_\mathrm{on-site}
	\sim JW^{-1}L^{-1}\,.
\end{align}
Thus we see that the "shape" or "geometry" of the anticrossings is invariant against disorder strength and system size as the ratio of the typical widths and typical spacings in both models is disorder and system size independent up to leading order (neglecting subleading $\sim\lambda V_0/L^2$ potential corrections, which anyway comes with a slightly small prefactor of $\sim0.2$), i.e. for $L\rightarrow\infty$ they neither disappear (limit of $\delta\lambda/\Delta\lambda\rightarrow0$) nor merge together (limit of $\delta\lambda/\Delta\lambda\rightarrow\infty$):
\begin{equation}
	\frac{\delta\lambda}{\Delta\lambda} \sim \mathrm{O}(1)\,.
\end{equation}
 The numerical verifications of our statements can be seen in Fig.~\ref{GOELZ} for the GOE ensemble showing the agreement of the gap and slope statistics with the analytical formulas~\cite{WilkAvCrossStat,WilkSingSpectrum} and with the inset also displaying the disorder independence of the number and of the gap of the anticrossings for the potential model. Moreover, in Fig.~\ref{GUSELZ} similar agreement is demonstrated for the LZ distributions in the case of the GUE and GSE ensembles with the insets verifying the size dependence of the number of the anticrossings for both models, the linear disorder strength scale of the number of the avoided crossings and the constant behavior of the gap for the on-site model.

\subsection{Universal single particle dynamics}

The universal "geometry" of the avoided level crossings also implies universal (disorder and size independent) transition rates and single particle dynamics in the case of slow quantum quenches, once proper velocity and time scales are chosen. For the sake of simplicity consider linear time-evolution in parameter space, $\lambda(t)=\lambda_i+vt$, with transition probabilities at the avoided crossings given by the celebrated Landau-Zener formula~\cite{LZ1,LZ2}:
\begin{equation}\label{eq:PLZ}
	P_\mathrm{LZ}=e^{-\frac{\pi}{2}\frac{\Delta^2_\mathrm{min}}{\gamma v}}.
\end{equation}
Introducing, in the case of slowly driven, near adiabatic processes, dimensionless time and velocity, measured in units determined by the gap and frequency of the avoided crossings $\tilde t=t/t_c,\,\,t_c=1/\langle\Delta_\mathrm{min}\rangle$, $\tilde v=v/v_c,\,\,t_cv_c=\Delta\lambda\Rightarrow v_c=\langle\Delta_\mathrm{min}\rangle\Delta\lambda$, leads to size and disorder independent Landau-Zener transition probabilities. To see this, consider the exponent of Eq.~\eqref{eq:PLZ} which implies a velocity scale of $v_c=\frac{\langle\Delta_\mathrm{min}\rangle^2}{\langle\gamma\rangle}$ in order to have universal transition strengths, which using the relation $\delta\lambda\sim\frac{\langle\Delta_\mathrm{min}\rangle}{\langle\gamma\rangle}$ leads to
\begin{equation}
	v_c=\frac{\langle\Delta_\mathrm{min}\rangle^2}{\langle\gamma\rangle}=\langle\Delta_\mathrm{min}\rangle\delta\lambda	
	\sim\langle\Delta_\mathrm{min}\rangle\Delta\lambda\,,
\end{equation}
where in the last step we used our knowledge about the fact that in both models, $\delta\lambda/\Delta\lambda$ is independent of both the system size and the disorder strength up to leading order. Moreover, for fixed dimensionless velocities and quench times we get the same number of the avoided crossings on average as well. Hence in slowly driven systems, where level-to-level transitions mostly happen at the anticrossings, on average the same number of such transitions happen with the same strength (up to leading order in the potential model) implying universal single particle dynamics.

\begin{figure*}[t]
\includegraphics[width=.47\textwidth]{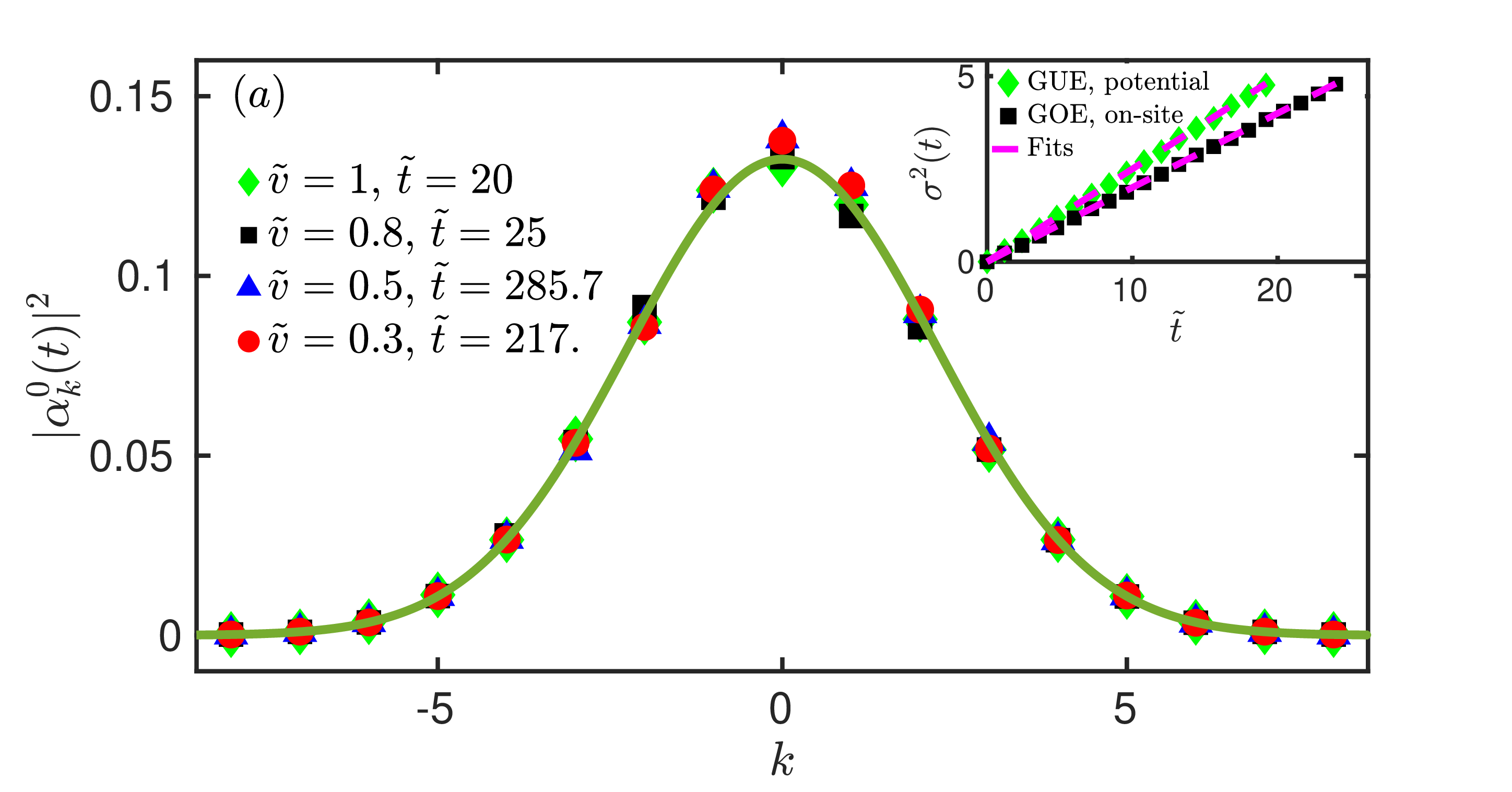}
\centering
\hfill
\includegraphics[width=.47\textwidth]{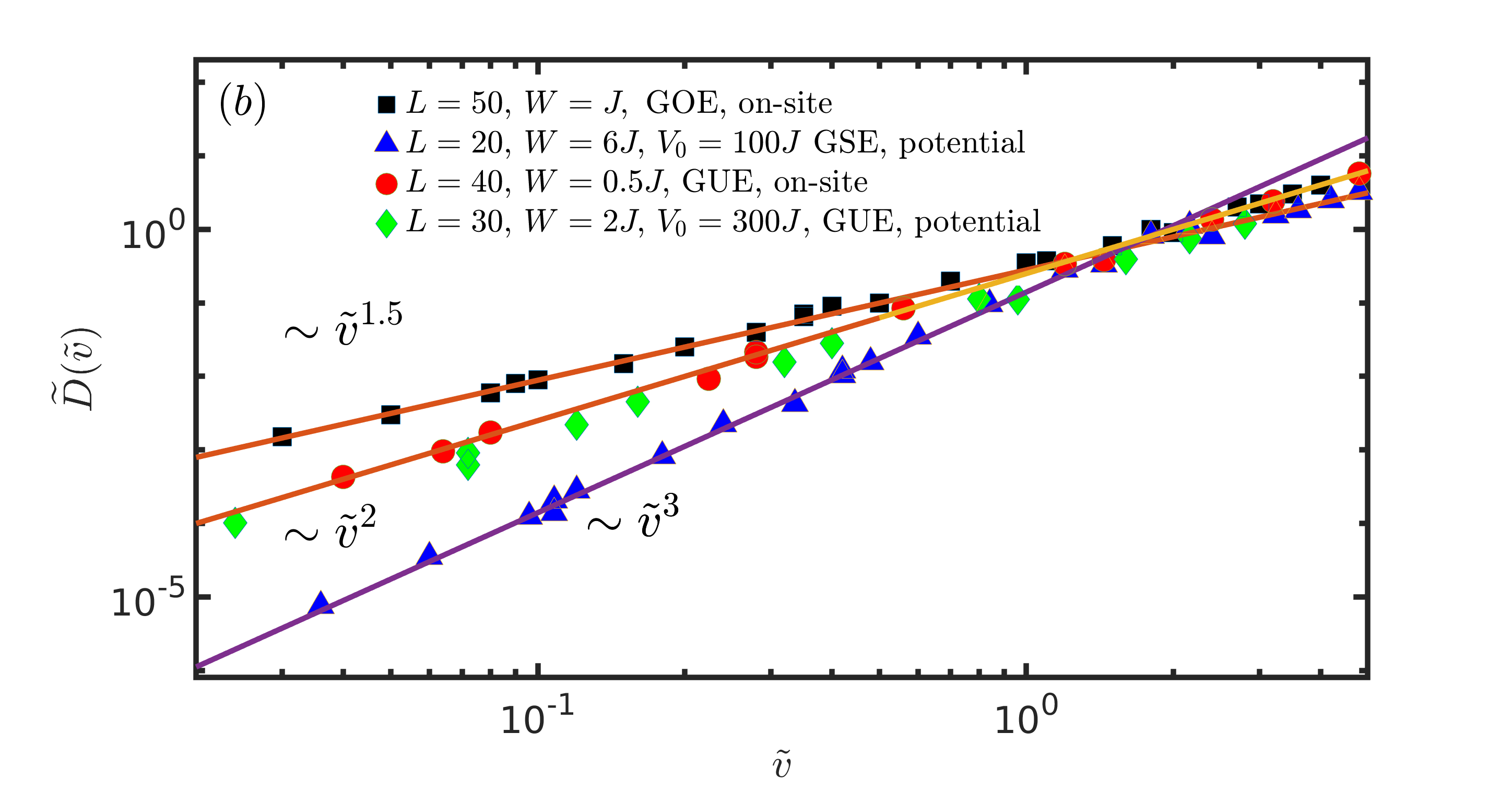}
\centering
\caption{(a): Verification of the universal single particle dynamics based on the universal distribution of the avoided level crossing parameters. Occupation profiles are plotted for all the three symmetry classes for velocities and quench times such that the occupation broadenings are the same. When scaling velocity and time as indicated in the main text these profiles become independent of the symmetry class, system size and disorder and potential strength and fall on top of each other on a universal Gaussian curve. Inset: Time-evolution of the variance of the occuaption profile for $L=40,\,W=0.5J$ for the GUE ensemble in the on-site model exhibiting linear dependence. (b): Velocity dependence of the diffusion coefficient, exhibiting a universal power law behavior with an anomalous dependence, $\sim\tilde v^{\beta/2+1}$ in the slow process limit with the slightly smaller values for the potential model. Numerical simulations were performed for various system sizes, disorder and potential strengths, for the three symmetry classes being identical on the two panels.}
\label{fig:DvUniv}
\end{figure*}
To gain a clearer description of the units, we formulate it also in terms of the disorder strength and system size, with the proper prefactors obtained numerically:
\begin{align}
	&v_\mathrm{pot}\sim L^{-4}J,\,\,\,\,\,\,\,\,\,\,\,\,\,\,\,\,\,\phantom{nnn}t_\mathrm{pot}\sim L^2J^{-1}\,,\\
	&v_\mathrm{on-site}\sim L^{-3}W^{-1}J^2,\,\,t_\mathrm{on-site}\sim L^2J^{-1}\,.
\end{align}
Next, we verify the above statements by numerically computing the time-evolution of the wave-function of a single fermion initially prepared in an eigenstate of $H(\lambda=0)$ with eigenenergy being closest to zero (implying the middle of the spectrum in the on-site model), denoted by $\boldsymbol\eta^0_0$ with $H(\lambda=0)\,\boldsymbol \eta^0_0=\varepsilon_0(0)\,\boldsymbol \eta^0_0$. For solving the time-dependent Schr\"odinger equation, $i\partial_t\boldsymbol\varphi^0(t)=H(t)\boldsymbol\varphi^0(t)$, we applied the adiabatic approach, i.e. we expanded the time-evolved wave-function in terms of the instantaneous eigenstates, $\boldsymbol\varphi^0(t)=\sum_{k=1}^{L^2}e^{-i\Phi_k(t)}\alpha^0_k(0)\boldsymbol\eta^0_t$ with $\Phi_k(t)=\int_0^t\mathrm dt^\prime\varepsilon_k\left(t^\prime\right)$ denoting the dynamical phase and transferred the differential equation to these expansion coefficients encoding the dynamics of the single fermion:
\begin{equation}
	\partial_t\alpha^0_k(t)=\sum_{k=1;\,k\neq l}^{L^2}A_{kl}(t)\alpha^0_l(t)
\end{equation}
with $A_{kl}(t)=-i\frac{\boldsymbol\eta^k_t\cdot \partial_tH(t)\cdot\boldsymbol\eta^l_t}{\varepsilon_k(t)-\varepsilon_l(t)}$. Finally, we implemented an RK4 routine to solve the above equation and obtain the expansion coefficients.
The dynamics is then perfectly characterized in terms of the occupation probabilities of the $k$th instantaneous eigenstates at time $t$ given by $\left\lvert\alpha^0_k(t)\right\rvert^2$ and the time evolution of their typical width.

In the case of slowly driven systems with $\tilde v\lesssim1$ the dynamics governed by the Landau-Zener transitions at the avoided level crossings and a classical Markovian energy space diffusion picture applies up to high accuracy~\cite{LZDiffusion,LZResponse,WilkDiffDissLZ}. Here both of the above quantitites exhibit universal behavior as predicted for fixed $\tilde v$ and $\tilde t$.
Moreover, the classical underlying picture immediately implies a universal Gaussian shape of the occupation numbers, $\left\lvert\alpha^0_k(t)\right\rvert^2\approx\frac{\exp(-k^2/4\widetilde D\tilde t)}{\sqrt{4\pi \widetilde D\tilde t}}$, and a linear time evolution of its variance, $\sigma^2(t)=\sum_{k=1}^{L^2}k^2\left\lvert\alpha^0_k(t)\right\rvert^2\approx2\widetilde D(\tilde v)\,\tilde t$ with $\widetilde D(\tilde v)$ being the universal, purely velocity dependent diffusion constant. We checked the validity of this universality for various values of $L,\,W$ and $V_0$ for both models and the three ensembles. In particular choosing such $\tilde v$ and $\tilde t$ that the broadening, $\sigma(t)$, of the occupation profile remains the same, occupation profiles fall on the same Gaussian curve as demonstrated in panel $(a)$ of Fig.~\ref{fig:DvUniv} with the inset verifying the linear time-dependence. In addition, we also plotted the numerically obtained values of the diffusion constant depending only on the dimensionless velocity (see Fig.~\ref{fig:DvUniv} panel $(b)$) and exhibiting in this diffusion regime an anolamous power-law behavior, $D(\tilde v)\sim\tilde v^{\beta/2+1}$ for $\tilde v\lesssim 1$, which is a direct consequence of the different strengths of level repulsion in the given ensembles. Note that the breakdown of the power-law curve starts around $\tilde v\approx1$, as predicted, and there is a small difference in the coefficients between the two models. For the derivation of the power-law behaviors, see Refs.~\cite{LZDiffusion,LZResponse,WilkDiffDissLZ} (although in these works units were defined with different coefficients).\\

\section{Conclusions}
In this work we first investigated the statistical properties of the dynamics of energy levels of disordered two-dimensional quantum dot models defined in~Eqs. \eqref{eq:onsite} and \eqref{eq:pot}. In the on-site model~\eqref{eq:onsite}, the same quench protocol was applied as in RMT, but with randomness only involved in the on-site energies with nearest neighbor hoppings, while in the potential model~\eqref{eq:pot} with fixed on-site random energies and hopping terms the motion of the energy levels was generated by a parabolic potential compressed (decompressed) in the $x\,(y)$ direction. First we considered the effect of the potential on the localization properties and the statistics of the velocity and curvature of energy levels. The localization effects, induced by the confining potential were captured by the average level spacing ratio. Next we determined how the variance of the level velocity and the curvature unit scaled with the system size and disorder strength and determined their statistics in the strongly localized regimes.
Moreover, we also provided numerical results on the threshold values in the two models below which their statistics is described by the RMT prediction up to a fixed precision.

In the third section we investigated the statistical properties of the avoided level crossings, playing important role in near adiabatic non-equilibrium processes. In a similar way, almost perfect agreement was found between the RMT results and the disordered models within the obtained regime determined by the threshold values of $L/\xi$ and $L/\xi_V$. Considering the size and disorder dependences of the LZ parameters we found agreement for the gap for both the two-dimensional models and the RMT results scaling with the inverse of the number of energy levels while insensitivity to disorder strength was observed.
The slope exhibited the same scalings as the velocity, i.e. independence of both the disorder strength and the system size for the potential model, while linear growth with the disorder and inverse scaling with the system size for the on-stie model. Finally, the average number of avoided crossings increased with the number of levels and the system size for the potential and on-site model, respectively, while insensitivity and linear growth was observed with respect to the disorder in the potential and on-site model, respectively.
Despite the different scalings, we concluded this discussion with the fact that the natural time and velocity units matched the ones predicted by size and disorder strength independent Landau-Zener transition rates, implying universal slowly driven dynamics. For verifying the latter claim we performed numerical simulations for the slowly driven case, with dimensionless velocity $\tilde v\lesssim1$, for the time-evolution of a single fermion initially prepared at zero energy. The results faithfully mirrored the predictions, that is the instantaneous occupation probabilities and the time-evolution of its broadening showed diffusion like behavior falling on a universal curve independently of the disorder and potential strength, system size and the symmetry class.

Our findings for the level dynamics in two-dimensional disordered systems can be extended in many natural ways. To say the least, localization properties and spacing statistics of the Floquet eigenstates and quasi energies in cyclic drivings or finite size scaling analysis of the localization length and critical disorder in the potential model for the GSE ensemble can also provide a fruitful perspective for future research. Further tests could also be performed verifying the agreement with RMT, for instance in terms of the fidelity susceptibility capturing directly the localization characteristics of the eigenstates~\cite{FidSusc1,FidSusc2}.

\begin{acknowledgments} I thank M\'arton Kormos, Izabella Lovas and Gergely Zar\'and for insightful discussions.  
his research was supported by the National Research, Development and Innovation Office - NKFIH through research grants Nos. K134983, K138606, and SNN139581, and within the Quantum National Laboratory of Hungary program (Project No. 2017-1.2.1-NKP- 2017-00001).
\end{acknowledgments}

\end{document}